\documentclass[a4paper,10pt]{article}
\pdfoutput=1 

\usepackage{jheppub} 

\usepackage[T1]{fontenc} 

\usepackage{mathrsfs,wasysym}
\usepackage{booktabs}

\newcommand{\sss}{\scriptscriptstyle}
\newcommand{\LQCD}{\Lambda_{\rm\sss QCD}}
\newcommand{\Sud}{\mathcal{S}}
\newcommand{\as}{\alpha_s}

\newcommand{\Ord}{\mathcal{O}}
\newcommand{\Lum}{\mathscr{L}}

\newcommand{\mh}{m_{\sss\rm H}}
\newcommand{\mt}{m_{\sss\rm t}}
\newcommand{\mbottom}{m_{\sss\rm b}}
\newcommand{\mcharm}{m_{\sss\rm c}}
\newcommand{\mz}{m_{\sss\rm Z}}
\newcommand{\muf}{\mu_{\sss\rm F}}
\newcommand{\mur}{\mu_{\sss\rm R}}

\newcommand{\pr}{$^\prime$}

\newcommand{\plus}[1]{\left(#1\right)_+}
\newcommand{\plusq}[1]{\left[#1\right]_+}
\newcommand{\D}{\mathcal{D}}

\newcommand{\Dh}{\hat{\mathcal{D}}}
\newcommand{\Dm}{\mathcal{D}^{\log}}

\renewcommand{\Re}{\mathrm{Re}\:}
\renewcommand{\Im}{\mathrm{Im}\:}
\let\originalleft\left
\let\originalright\right
\renewcommand{\left}{\mathopen{}\mathclose\bgroup\originalleft}
\renewcommand{\right}{\aftergroup\egroup\originalright}

\newcommand{\CNsoft} {C_{N\text{-soft}}}

\newcommand{\AP}[1]{\mathcal{AP}_{#1}}

\def\beq{\begin{equation}}  
\def\eeq{\end{equation}}
\def\({\left(}
\def\){\right)}
\def\[{\left[}
\def\]{\right]}
\allowdisplaybreaks

\title{\boldmath Resummed Higgs cross section at N$^3$LL}

\author[a]{Marco Bonvini}
\affiliation[a]{Deutsches Elektronen-Synchroton, DESY, Notkestra{\ss}e 85, D-22603 Hamburg, Germany}
\author[b]{and Simone Marzani}
\affiliation[b]{Institute for Particle Physics Phenomenology, Durham University, South Road, Durham DH1 3LE, England}
\preprint{
\begin{flushright}
DESY 14-075, \\ 
DCPT/14/94, \\ 
IPPP/14/47
\end{flushright}
}

\emailAdd{marco.bonvini@desy.de}
\emailAdd{simone.marzani@durham.ac.uk}

\abstract{

  We present accurate predictions for the inclusive production of a
  Higgs boson in proton-proton collisions, via gluon-gluon fusion. Our
  calculation includes next-to-next-to-leading order (NNLO)
  corrections in perturbative QCD, as well as the resummation of
  threshold-enhanced contributions to next-to-next-to-next-to-leading
  logarithmic (N$^3$LL) accuracy, with the inclusion of the recently-determined three-loop constant coefficient (sometimes referred to as N$^3$LL\pr\ accuracy).

  Our result correctly accounts for finite top, bottom and charm
  masses at leading order (LO) and next-to-leading order (NLO), and
  includes top mass dependence at NNLO.
  At the resummed level the dependence on top, bottom and charm mass is accounted for at NLL,
  while only the top mass at NNLL.
  The all-order calculation is improved by a suitable choice
  of the soft terms, dictated by analyticity conditions and by the
  inclusion of subleading corrections of collinear origin, which
  improve the accuracy of the resummation away from the threshold
  region.

  We present results for different collider energies and we study
  perturbative uncertainties by varying renormalization and
  factorization scales.
  We find that, at current LHC energies, the resummation corrects the NNLO result by as much as 20\% at $\mur=\muf=\mh$, while the correction is much smaller, 5.5\%, at $\mur=\muf=\mh /\,2$. 
  While the central value of NNLO+N$^3$LL result depends very mildly on the scale choice, we argue that a more reliable estimate of the theoretical uncertainty is found if the perturbative scales are canonically varied about $\mh$.

}

\begin{document}

\maketitle
\flushbottom

\section{Introduction}

The resummation of soft-gluon (or threshold) logarithms in QCD plays
an important role in precision phenomenology at hadron colliders, and
in particularly at the LHC.  Examples include Higgs boson production
in gluon fusion, e.g.~\cite{Catani:2003zt}, top-pair production,
e.g.~\cite{Cacciari:2011hy} and supersymmetric particles,
e.g.~\cite{Beenakker:2014sma}. Soft gluon resummation improves the
accuracy of the predicted cross section, leading, for instance, to a
reduced scale dependence.

This is particularly important in the case of Higgs production in
gluon-gluon fusion. QCD corrections are fully known up to
next-to-next-to-leading order (NNLO) accuracy~\cite{higgsuptoNNLO,
  higgsuptoNNLO-finite-mt} and N$^3$LO calculations are
underway~\cite{NNNLO}. Very recently, the first term in the soft
expansion of the full N$^3$LO cross section has been
obtained~\cite{Anastasiou:2014vaa}.
The perturbative behavior of this series is very poor and thus
logarithmically enhanced soft terms, predicted to all orders by soft-gluon
resummation, provide a powerful tool to include (and check) higher
order terms in the series.

Strictly speaking, soft-gluon resummation is \emph{needed} when the partonic subprocess
is close to threshold: being $M$ the mass of the tagged final state (the Higgs boson mass, for instance)
and $\sqrt{\hat s}$ the center-of-mass energy of the partonic subsystem, in the
limit $z=M^2/\hat s \to 1$ the QCD perturbative expansion of the partonic cross section
is unstable and the resummation of the entire series is necessary.
Whether this is the case in the computation of the physical hadron-level cross section
depends both on hadron-level kinematics and on the shape of parton distribution functions (PDFs),
since the physical cross section is a convolution of the partonic cross section and PDFs.

In most cases, and particularly for inclusive observables at the LHC, the partonic region
$z\to 1$ for which perturbativity is lost gives only a moderate, often negligible,
contribution to the physical cross section~\cite{Bonvini:2012an}.
In these cases, soft-gluon resummation is no longer \emph{needed};
however, it might still be \emph{advisable}.
Indeed, an intermediate range of values of $z$ for which
the soft terms approximate well the full partonic cross section usually exists~\cite{Bonvini:2012an,Ball:2013bra};
although in this region the series is behaving in a perturbative way,
inclusion of higher order terms from soft-gluon resummation leads to a
more accurate and stable prediction for the partonic cross section.
When this intermediate range dominates the physical cross section,
soft-gluon resummation provides then a powerful way of including (the dominant part of)
higher order terms in the perturbative expansion.

The ability of all-order calculations to capture the region of intermediate $z$ strongly depends on
the actual form of the soft terms that are being resummed~\cite{Bonvini:2012an,Ball:2013bra}.
Indeed, while the soft limit determines the large-$z$
(or large-$N$, being $N$ the conjugate variable of $z$ upon Mellin transformation)
behavior of the soft terms, it does not fix their functional form.
Traditionally, the $N$-space resummation of the soft terms is organised in terms of powers of $\log N$ and constants. 
However, using analyticity arguments~\cite{Ball:2013bra}, we arrived at the conclusion that this choice
is not optimal in several respects, chiefly because powers of $\log N$ exhibit a branch cut at finite $N$,
in contrast to the pole structure of fixed-order coefficient functions.
A form of the resummation that respects these analyticity properties is advisable,
in particular if we aim to capture the dominant behavior in the region of intermediate $N$, i.e.\ intermediate $z$.

Following our previous studies~\cite{Ball:2013bra,Bonvini:2014jma}, we
consider a functional form for the soft terms that respects the
analyticity properties of fixed-order results.  We improve on that
work by implementing this formalism in an all-order resummation
formula and by computing the Higgs production cross section at
NNLO+N$^3$LL, for different collider energies.
Our result includes all the information from the N$^3$LO soft-virtual calculation of Ref.~\cite{Anastasiou:2014vaa}, and reproduces to order $\as^3$
the soft part of the N$^3$LO approximate prediction of Ref.~\cite{Bonvini:2014jma}.
We also study a
different form of soft terms, which has the advantage of respecting
the aforementioned analyticity conditions, while having, at the same time,
a fast numerical implementation.

Finally, we note that a form of threshold resummation with the correct singularity
structure at finite $N$ is a necessary step towards the construction of a double-resummed
cross section in which threshold and high-energy (BFKL) logarithms are simultaneously
accounted for to all orders.

\section{Soft-gluon resummation}

\subsection{Generalities}

A physical (hadron-level) inclusive cross section at hadron colliders can be written in the factorized form
\beq\label{eq:xs}
\sigma(\tau,M^2) = \tau
\sum_{ij}\int_\tau^1 \frac{dz}{z}\,\Lum_{ij}\(\frac{\tau}{z},\muf^2\)
\frac1z \hat\sigma_{ij}\(z, M^2, \as(\mur^2),\frac{M^2}{\muf^2},\frac{M^2}{\mur^2}\),
\qquad
\tau=\frac{M^2}{s},
\eeq
where $\Lum_{ij}(z,\mu^2)$ is a parton luminosity
\beq\label{eq:lum}
\Lum_{ij}(z,\mu^2) = \int_z^1 \frac{dx}x\, f_i\(\frac zx,\mu^2\) f_j(x,\mu^2),
\eeq
and $i,j$ run over all parton flavours.
Without loss of generality, we can suppress the flavour indices
and concentrate on the dominant channels for soft resummation ($gg$ for Higgs).
For ease of notation, we also suppress factorization scale $\muf$ and renormalization scale $\mur$ dependence.
The partonic cross section $\hat\sigma$ is related to the so-called dimensionless coefficient function $C$ by
\beq\label{eq:partonic_xs}
\hat\sigma(z,M^2) = z\,\sigma_0(M^2) \,C(z,\as),
\eeq
where $\sigma_0$ is the leading order (LO) partonic cross section, so that
the coefficient function is normalized to $\delta(1-z)$ at leading order:
\beq\label{eq:cf_exp}
C(z,\as) = \delta(1-z) + \as C^{(1)}(z) + \as^2 C^{(2)}(z) + \ldots,
\eeq
and $z=M^2/\hat s$ is the variable already mentioned in the introduction.
In terms of this coefficient function the cross section Eq.~\eqref{eq:xs} reads
\beq
\sigma(\tau,M^2) = \tau\,\sigma_0(M^2)\int_\tau^1 \frac{dz}{z}\,\Lum\(\frac{\tau}{z}\) C\(z,\as(M^2)\),
\eeq
which has the form of a Mellin convolution, and factorizes in Mellin space
\beq\label{eq:xs_mell}
\sigma(N,M^2)=\int_0^1d\tau\,\tau^{N-1}\frac{\sigma(\tau,M^2)}\tau =\sigma_0(M^2)\, \Lum(N)\, C\(N,\as(M^2)\).
\eeq
Note that we have used the same symbols, with different arguments, for a function and its Mellin transform; note also that, for convenience, we have indicated with
$\sigma(N,M^2)$ the Mellin transform of $\sigma(\tau,M^2)/\tau$.

Soft-gluon resummation is generically performed in $N$-space,
where the multiple gluon emission phase-space factorizes.
The $N$-space resummed coefficient function (for Higgs and Drell-Yan production)
has the form~\cite{resummation} 
\begin{align}
\label{eq:Cres}
C_{\rm res}\(N,\as\)&=\bar g_0\(\as\) \exp\bar\Sud(\as,N),\\
\bar\Sud(\as,N)
&= \int_0^1dz\,z^{N-1} \plusq{\frac1{1-z}
\( \int_{\muf^2}^{M^2(1-z)^2} \frac{d\mu^2}{\mu^2} 2A\(\as(\mu^2)\) + D\(\as([1-z]^2M^2)\) \)} \nonumber\\
&= \int_0^1dz\, \frac{z^{N-1}-1}{1-z}
\( \int_{\muf^2}^{M^2(1-z)^2} \frac{d\mu^2}{\mu^2} 2A\(\as(\mu^2)\) + D\(\as([1-z]^2M^2)\) \),\label{eq:Shat}\\
\bar g_0(\as) &= 1+\sum_{k=1}^\infty \bar g_{0,k} \as^k,\\
A(\as)&=\sum_{k=1}^\infty A_k\as^k, \qquad
D(\as)=\sum_{k=1}^\infty D_k\as^k,\label{eq:A,D}
\end{align}
where $\as=\as(\mur^2)$, and $\bar g_0(\as)$ does not depend on $N$, but depends implicitly
on $\muf/M$ and $\mur/M$. The function $A(\as)$
(also called cusp anomalous dimension $\Gamma_{\rm cusp}$) is the numerator
of the divergent part of the relevant\footnote{Depending on the considered process,
the relevant splitting function can be $P_{gg}$ (Higgs) or $P_{qq}$ (Drell-Yan); the corresponding $A_g(\as)$
and $A_q(\as)$ functions are simply related by a color factor: $C_F A_g(\as) = C_A A_q(\as)$.
The same color-charge relation holds for $D(\as)$: $C_F D_{\rm Higgs}(\as) = C_A D_{\text{Drell-Yan}}(\as)$.}
diagonal Altarelli-Parisi splitting function,
\beq \label{eq:dglap-soft}
P(z,\as) = \frac{A(\as)}{\plus{1-z}} + B(\as)\delta(1-z) + \Ord\( \(1-z\)^0 \),
\eeq
and $D(\as)$ is a process-dependent function.
\begin{table}[t]
\begin{center}
\begin{tabular}{llcccrl}
  Notation* & Notation\pr & $A(\as)$ & $D(\as)$ & $\bar g_0(\as)$ & $C_{\rm res} \ni \as^n L^k\quad \forall n$ & adopted in\\
  \midrule
  LL     &   LL     & 1-loop & --- & tree-level & $k= 2n$ &\\
  \addlinespace[0.8\defaultaddspace]
  NLL*   &   NLL    & 2-loop & 1-loop & tree-level & $2n-1\le k\le 2n$ &\\
  NLL    &   NLL\pr   & 2-loop & 1-loop & 1-loop & $2n-2\le k\le 2n$ &\\
  \addlinespace[0.8\defaultaddspace]
  NNLL*  &   NNLL   & 3-loop & 2-loop & 1-loop & $2n-3\le k\le 2n$ &\\
  NNLL   &   NNLL\pr  & 3-loop & 2-loop & 2-loop & $2n-4\le k\le 2n$ & Refs.~\cite{Catani:2003zt,deFlorian:2012yg}\\
  \addlinespace[0.8\defaultaddspace]
  N$^3$LL* &   N$^3$LL  & 4-loop & 3-loop & 2-loop & $2n-5\le k\le 2n$ & Ref.~\cite{Ahrens:2008nc}\\
  N$^3$LL  &   N$^3$LL\pr & 4-loop & 3-loop & 3-loop & $2n-6\le k\le 2n$ & this work
\end{tabular}
\caption{Orders of logarithmic approximations and accuracy of the
  predicted logarithms $L=\log N$. See Ref.~\cite{Bonvini:2013td}.
  Note that the four-loop contribution to $A(\as)$ is yet unknown
  and in our study we take a Pad\'e approximation~\cite{Moch:2005ba}.}
\label{tab:count}
\end{center}
\end{table}
A given logarithmic accuracy is obtained including the functions
$A(\as)$, $D(\as)$ and $\bar g_0(\as)$ up to a given order in Eq.~\eqref{eq:Cres},
according to Table~\ref{tab:count}. 

Table~\ref{tab:count} shows two notations for the counting of logarithms,
usually adopted in different contexts.
In particular, in the two notation what is called N$^k$LL for $k>0$ represents
two different accuracies, and therefore can lead to some confusion.
In what we call Notation\pr, the N$^k$LL accuracy without decoration
corresponds to a logarithmic counting on $\log C_{\rm res}$, where the inclusion of an additional
order in $\bar g_0(\as)$, corresponding to N$^k$LL\pr, does not increase the formal accuracy.
However, as shown explicitly in the table, N$^k$LL\pr\ predicts an additional term to all orders in
the tower of logarithms contained in $C_{\rm res}$.
In practice, when the process is not very close to the physical threshold (i.e., $\tau$ is not close to $1$),
as in all relevant cases, the additional logarithm included at N$^k$LL\pr\ improves the actual accuracy
of the result.
Therefore, in a wide literature, the N$^k$LL\pr\ is simply called N$^k$LL, as shown in what we call Notation*,
where the lower accuracy is denoted with a *, although this notation is not widespread.
In what follows, we will refer to a logarithmic accuracy according to Notation*.

The three-loop coefficients of $A(\as)$ and $D(\as)$ have been known
for while (see for instance Refs.~\cite{Moch:2005ba, D3}), while the $\Ord
\(\as^3 \)$ contribution to $\bar g_0(\as)$ has been recently computed
in the infinite top-mass limit~\cite{Anastasiou:2014vaa}. The function
$A(\as)$, however, is needed at four loops in order to achieve full
N$^3$LL accuracy. This contribution is yet unknown; however, a Pad\'e
estimate~\cite{Moch:2005ba} can suggest the size of its value, and a
numerical analysis shows that its impact in a resummed result is
essentially negligible.
Furthermore, we note that, even without this contribution, the expansion of the N$^3$LL\pr\ resummation to third order in the strong coupling completely reproduces all the soft and constant contributions up to N$^3$LO, i.e.\  the four loop coefficient of $A(\as)$ only enters at order $\as^4$.

\subsection{Large $N$ limit --- $N$-soft}\label{sec:nsoft}

The Mellin transform in Eq.~\eqref{eq:Shat} is ill defined, because $z$ ranges from $0$ to $1$,
forcing the argument of $\as$ to be arbitrarily small, therefore crossing the Landau pole.
However, the integral can be made convergent by using the explicit solution for the running coupling
\beq
\as(\mu^2) = \frac{\as(\mur^2)}{X} - \frac{\as^2(\mur^2)}{X^2}\frac{\beta_1}{\beta_0} \log X +\ldots,
\qquad X=1+\beta_0 \as(\mur^2) \log\frac{\mu^2}{\mur^2}
\eeq
to any finite order.
In this way, $\bar\Sud$ can be formally written as
\begin{align}
\bar\Sud(\as,N) &= \int_0^1dz\,z^{N-1} \bar\Sud(\as,z) = \sum_{n=1}^\infty \as^n \sum_{k=0}^n b_{n,k} \D_k(N), \label{eq:ShatMellin}\\
\bar\Sud(\as,z) &= \sum_{n=1}^\infty \as^n \sum_{k=0}^n b_{n,k} \D_k(z) \label{eq:Shatser}
\end{align}
where
\beq\label{eq:Dk}
\D_k(z) = \plus{\frac{\log^k(1-z)}{1-z}}
\eeq
are the usual plus-distributions and
\beq\label{eq:Dkmell}
\D_k(N)= \int_0^1 d z\, z^{N-1} \D_k(z)=
 \frac{1}{k+1}\sum_{j=0}^{k+1}\binom{k+1}{j}\, \Gamma^{(j)}(1)\,
\[\frac{d^{k+1-j}}{d\xi^{k+1-j}} \frac{\Gamma(N)}{\Gamma(N+\xi)}\]_{\xi=0}
\eeq
their Mellin transform.

The Mellin transform in Eq.~\eqref{eq:ShatMellin} is usually computed, to any finite logarithmic accuracy, in the large-$N$ limit, leading to an expression of the form
\begin{align}
\label{eq:Cres2}
\CNsoft(N,\as) &=g_0(\as) \exp\Sud(\as,\log N),\\
\Sud(\as,\log N) &= \left[\frac{1}{\as} g_1(\as\log N)+ g_2(\as\log N)+ \as g_3(\as\log N)+ \as^2 g_4(\as\log N)+\dots\],\\
g_0(\as) &= 1+\sum_{k=1}^\infty g_{0,k} \as^k,\\
g_i(\lambda)&=\sum_{k=1}^\infty g_{i,k}\lambda^k, \qquad g_{1,1}=0,
\end{align}
where we have introduced a new notation ($N$-soft) for the resummed coefficient function
to stress the fact that the large-$N$ limit has been taken. Note that
\beq \label{eq:res-vs-Nsoft}
\CNsoft(N,\as)=C_\text{res}(N,\as) \[1+\Ord\(\frac{1}{N}\)\].
\eeq

The functions $g_i$, $i=1,2,3,4$ can be found explicitly for many processes in Ref.~\cite{Moch:2005ba}.
A given N$^k$LL accuracy is obtained from Eq.~\eqref{eq:Cres2} including
$g_i$ up to $i=k+1$, and $g_0(\as)$ up to the same order as $\bar g_0(\as)$, see Table~\ref{tab:count}.
The function $\Sud(\as,\log N)$ can be written as
\beq\label{eq:Sudakov_expansion}
\Sud(\as,\log N) = \sum_{n=1}^\infty \as^n \sum_{k=0}^n b_{n,k} \Dm_k(N),
\eeq
where the coefficients $b_{n,k}$ are the same as in Eq.~\eqref{eq:ShatMellin},
and the functions $\Dm_k(N)$ are the large-$N$ limit of $\D_k(N)$
expressed in terms of $\log N$, and neglecting constant terms and terms suppressed by powers of $1/N$:
\beq \label{eq:DklogN}
\Dm_k(N)
=\frac{1}{k+1}\sum_{j=0}^k\binom{k+1}{j}\, \Gamma^{(j)}(1)\,
\log^{k+1-j}\frac{1}{N}.
\eeq
Here $\Gamma^{(j)}(x)$ is the $j$-th derivative of the Euler Gamma function.
For completeness, we also report the functional form of the momentum space conjugates of the $\Dm_k(N)$ functions: 
\beq\label{dmdef}
\Dm_k(z)= \plus{\frac{\log^k\log\frac1z}{\log\frac1z}}, \quad \text{with} \quad \Dm_k(N)=\int_0^1 d z\, z^{N-1} \Dm_k(z)\,.
\eeq
The relation between the function $g_0$ and the constant that multiplies the resummed exponent in Eq.~\eqref{eq:Cres}, namely $\bar g_0$, is
\beq \label{eq:g0bardef}
g_0(\as)= \bar g_0(\as)  \exp\[\sum_{n=1}^\infty \as^n \sum_{k=0}^{n}  b_{n,k}d_k\],
\eeq
with
\beq \label{eq:dk_def}
d_k=\lim_{N\to\infty}\[\D_k(N)-\Dm_k(N)\] = \frac{\Gamma^{(k+1)}(1)}{k+1},
\eeq
(for further details, see Ref.~\cite{Ball:2013bra}).

\subsection{Prescriptions for the Landau pole}
\label{sec:prescriptions}

The resummed coefficient function $\CNsoft$, Eq.~\eqref{eq:Cres2}, although well defined in $N$ space, cannot be directly used for computing the corresponding hadron-level cross section
because its inverse Mellin transform does not exist.
Indeed, the functions $g_i(\as\log N)$ have a branch-cut for real
$N>N_L=\exp\frac1{2\beta_0\as}$ originating from the Landau pole of the running coupling,
while Mellin transformation has a convergence abscissa~\cite{Catani:1996yz}.
On the other hand, if $\CNsoft$ is expanded in powers of $\as$, the inverse Mellin transform exists
to any finite order, but the resulting series is divergent~\cite{Forte:2006mi}.
Therefore, a prescription is needed to compute physical observables from Eq.~\eqref{eq:Cres2}.

The most used prescription is the so called Minimal Prescription (MP), proposed long ago~\cite{Catani:1996yz}.
It consists on a simple modification of the Mellin inversion integral,
and has the advantage of having a fast numerical implementation.
More details are given in App.~\ref{app:MP}.
More recently, another prescription based on a Borel summation of the divergent series of the
order-by-order inverse Mellin transform of $\CNsoft$ has been first proposed in Ref.~\cite{Forte:2006mi},
and refined and extended in Refs.~\cite{Abbate:2007qv,Bonvini:2008ei,Bonvini:2010tp,Bonvini:2012sh}.
This prescription, called Borel Prescription (BP), is typically slower but more flexible.
More details are given in App.~\ref{app:BP}.

It turns out that the numerical difference between the two prescriptions is small, being \emph{totally} negligible for Higgs phenomenology~\cite{Bonvini:2012sh}.
The reason is that, for the kinematic configurations typical of high-energy colliders, i.e.\ $\tau\ll1$,
the series is behaving in a perturbative way.
Hence, the all-order nature of the series does not play any role, and,
a fortiori, the way the divergence of the series is dealt with is immaterial.
We believe that expanding $\CNsoft$ to a sufficiently large, but \emph{finite}, order in $\as$
and inverting exactly would lead to a result virtually identical to the all-order MP or BP results.

One reason why the effective equivalence of MP and BP does not appear clearly
in the literature is the fact that, within the BP, the form of the soft terms
that are resummed can be easily modified, and this has been always done for phenomenological
application, thereby giving a result which differs from the MP one.
A discussion on the form of the soft terms will be performed in Sect.~\ref{sec:soft-analytic}.
Here, we just want to mention that, for practical applications, the choice of
the prescription will be mainly dictated by its flexibility and numerical efficiency.

Details on the practical implementation of the prescriptions, as well as details
on a new version of the BP acting directly on the Sudakov exponent $\Sud(\as,\log N)$,
are given in App.~\ref{app:prescriptions}.

\section{Soft terms and analyticity conditions}
\label{sec:soft-analytic}

Soft gluon resummation fixes the coefficients $b_{n,k}$ of the soft
terms in Eq.~\eqref{eq:Sudakov_expansion}, however it does not fix the
functional form of these contributions.  In particular, choices that
at large $N$ only differ by terms suppressed by factors of $1/N$ are
equally acceptable.

It has been pointed out~\cite{Kramer:1996iq,Ball:2013bra} that the actual form
of the soft terms is very important, and different choices would lead to very different
accuracies, in particular when the considered process is far from the physical threshold.

In our previous analysis~\cite{Ball:2013bra}, we studied the
analyticity properties of coefficient functions in $N$ space and we
arrived at an optimal choice of the soft terms. We used this improved
large-$N$ behavior, together with the knowledge of the rightmost
singularity at finite $N$ from high-energy resummation to compute an
approximate expression for the N$^3$LO Higgs production
cross section~\cite{Ball:2013bra, Bonvini:2014jma}.

In this section, we briefly review the two main theoretical
ingredients that go into the construction of our resummation formula,
namely a choice of soft terms that respects the singularity structure
of coefficient functions and the improvement related to the inclusion
of collinear
contributions. 

\subsection{Functional form of the soft terms}\label{sec:func-form}

The $N$-soft resummed exponent Eq.~\eqref{eq:Sudakov_expansion} is an
infinite sum of contributions each of which has a logarithmic branch
cut starting at $N=0$, which is not compatible with the known
singularity structure of coefficient functions. However, this problem
is an artefact of the large $N$ approximation which we have employed
in going from $\bar \Sud$, Eq.~\eqref{eq:ShatMellin}, to $\Sud$,
Eq.~\eqref{eq:Sudakov_expansion}. Indeed, the resummed exponent $\bar
\Sud$ is written as an infinite sum of $\D_k(N)$ functions, whose
singularity structure is compatible with the one of fixed-order
calculations.

Furthermore, as we discussed at length in Ref.~\cite{Ball:2013bra}, we
can improve on the use of $\D_k(N)$ by noticing that the correct
kinematic limit of the $\mu^2$ integration in Eq.~\eqref{eq:Shat} is
actually $M^2\frac{(1-z)^2}{z}$.
This consideration leads to the following choice for the soft terms in momentum space
\beq\label{eq:Dhat-z}
\Dh_k(z)=\D_k(z)+\frac{\log^k\frac{1-z}{\sqrt{z}}}{1-z}-\frac{\log^k(1-z)}{1-z}
= \[\frac{d^k}{d\xi^k} \(z^{-\xi/2}\plusq{(1-z)^{\xi-1}}\)\]_{\xi=0},
\eeq
from which we can easily compute the functions $\Dh_k(N)$, which enter our resummation formula:
\beq \label{eq:Dhat}
\Dh_k(N)= \int_0^1 d z\, z^{N-1} \Dh_k(z)=
 \frac{1}{k+1}\sum_{j=0}^{k+1}\binom{k+1}{j}\, \Gamma^{(j)}(1)\,
\[\frac{d^{k+1-j}}{d\xi^{k+1-j}} \frac{\Gamma(N-\xi/2)}{\Gamma(N+\xi/2)}\]_{\xi=0}.
\eeq
Note that in Eq.~\eqref{eq:Dhat-z} we have chosen to apply the plus prescription only to the
first term, singular in $z=1$, which is the natural choice in
fixed order calculations. 
In this way, $\Dh_k(N)$ differs from
$\D_k(N)$ only by terms vanishing at large $N$:
\beq \label{eq:DminusDhat}
\lim_{N\to\infty} \[\Dh_k(N)-\D_k(N)\] = 0.
\eeq
Adopting this form for the soft terms in all the terms generated in Eq.~\eqref{eq:Shat}
(hence also those generated by the $D(\as)$ term), we arrive at the expression
\beq\label{eq:ShatMellin2}
\bar\Sud(\as,N) \to \sum_{n=1}^\infty \as^n \sum_{k=0}^n b_{n,k} \Dh_k(N),
\eeq
which represents the first improved version of Eq.~\eqref{eq:ShatMellin}.
This is not yet our final formula, as we are going to discuss.

\subsection{Altarelli-Parisi contributions}\label{sec:altarelli-parisi}

We have already discussed the origin of the coefficient $A(\as)$ in
the resummation formula Eq.~\eqref{eq:Shat}: it is the coefficient of the
soft-enhanced part of the relevant Altarelli-Parisi splitting function
Eq.~\eqref{eq:dglap-soft}, where $P=P_{gg}$ in the case of Higgs
production. In resummed calculation, the coefficient $B(\as)$ in
Eq.~\eqref{eq:dglap-soft} is also retained at the appropriate accuracy
because it corresponds to a constant term in $N$-space,
while contributions that vanish as $z \to 1$ are usually neglected in
Eq.~\eqref{eq:Shat}. However, an important class of these subleading corrections
can be taken into account to all orders, essentially
because the full leading order anomalous dimension
exponentiates~\cite{Kramer:1996iq, Contopanagos:1996nh}.

However, the $1/z$ pole present in the LO gluon-gluon splitting
function would introduce spurious singularities in the resummed
coefficient function at $N=1$~\cite{Ball:2013bra}.
Nevertheless, the expansion of $(1-z) P_{gg}^{(0)}(z)$ (with $P_{gg}^{(0)}(z)$ being the LO
gluon-gluon Altarelli-Parisi splitting function) in powers of $1-z$
to any finite order is not singular in $z=0$, and therefore does not affect the singularity structure around $N=1$.
The expansion up to second order reads
\begin{align}
(1-z) P_{gg}^{(0)}(z)
&= A_1\[1-(1-z) + 2(1-z)^2 +\Ord\((1-z)^3\)\] \nonumber\\
&= A_1 \[2-3z+2z^2\] +\Ord\((1-z)^3\),  \label{eq:pgg-exp}
\end{align}
where $A_1=C_A/\pi$ is the first coefficient in the expansion of $A(\as)$, Eq.~\eqref{eq:A,D}.
Note that the third order term in the expansion is accidentally zero.
Upon Mellin transformation, multiplication by the factor $2-3z+2z^2$ results into the shift
\beq \label{eq:AP-operator}
\[2-3z+2z^2\]f(z) \overset{\text{Mellin}}{\longrightarrow} \AP2\[f(N)\]\equiv 2 f(N)-3 f(N+1)+2 f(N+2),
\eeq
where we have introduce the Altarelli-Parisi ($\AP2$) operator to second order for future convenience.

The $A_1$ term in Eq.~\eqref{eq:Shat} is responsible for the tower of LL terms,
namely terms $\as^n \D_{2n-1}(z)$ to all orders $n$ in $C(z,\as)$.
When $A_1$ is replaced with the expansion Eq.~\eqref{eq:pgg-exp}, logarithmic
terms suppressed by powers of $(1-z)$ are generated to all orders.
The towers of LL suppressed logarithms, namely terms of the form
$(1-z)^{k-1} \as^n \log^{2n-1}(1-z)$ in $C(z,\as)$,
with $k$ running from one up to the order of the expansion of $(1-z) P_{gg}^{(0)}(z)$,
are correctly predicted to all orders~\cite{Catani:2001ic}.
However, the simple inclusion of additional information from Altarelli-Parisi splitting functions
is not enough to predict terms beyond these LL towers.

Nevertheless, we have shown in Ref.~\cite{Ball:2013bra} that including the expansion of $(1-z) P_{gg}^{(0)}(z)$ up
to second order for all the terms in the Sudakov exponent Eq.~\eqref{eq:Shatser}
leads to an approximation of the exact fixed-order terms which is very good in a wide
range of $N$ values, down to values where high-energy terms (not considered in this work)
start being relevant.
This is achieved when the soft terms introduced in Sect.~\ref{sec:func-form} are used.
Therefore, the inclusion of such second order expansion for all the terms in Eq.~\eqref{eq:Shat} amounts to replacing
in Eq.~\eqref{eq:ShatMellin2}
\beq\label{eq:Dlog->Dhat_N2}
\Dh_k(N) \to \AP2 \[\Dh_k(N)\]=2\Dh_k(N) -3\Dh_k(N+1) +2\Dh_k(N+2).
\eeq
Note that the inclusion of terms of order $(1-z)^4$ and higher
in the expansion of $(1-z) P_{gg}^{(0)}(z) $ does not affect the results significantly~\cite{Ball:2013bra}. 
At resummed level, the operator $\AP2$ in Eq.~\eqref{eq:Dlog->Dhat_N2} can be applied directly
to the Sudakov exponent.

\subsection{$A$-soft$_2$}\label{sec:A2-soft}

We are now ready to present our resummed expression. We name it\footnote{This choice was called simply soft$_2$ in Ref.~\cite{Ball:2013bra}, where an average of soft$_1$ and soft$_2$ was used.}
$A$-soft$_2$, because it takes into account the analyticity properties
discussed in Sect.~\ref{sec:func-form} and it includes the first two
terms of the $(1-z)$ expansion of the LO splitting function, as
detailed in Sect.~\ref{sec:altarelli-parisi}:
\begin{align}\label{eq:A-soft2}
C_{A\text{-soft}_2}(N,\as)
&= \bar g_0(\as) \,\exp \sum_{n=1}^\infty \as^n \sum_{k=0}^{n}  b_{n,k}\,
\AP2\[\Dh_k(N)\].
\end{align}
Note that because the difference between the functions $\Dh_k(N)$ and
$\D(N)$ vanishes at large $N$, Eq.~\eqref{eq:DminusDhat}, the
constant term $\bar g_0$ in Eq.~\eqref{eq:A-soft2} coincides with the
one in Eq.~\eqref{eq:Cres}.
As in the case of $N$-soft resummation previously discussed, the
series that defines $C_{A\text{-soft}_2}$ in Eq.~\eqref{eq:A-soft2} is
divergent. This series is summed using the Borel Prescription\footnote{The MP cannot be used in this case,
because it is not flexible enough to reproduce the desired soft terms.} detailed
in App.~\ref{app:BP}, which leads to the following result
\begin{multline}
C_{A\text{-soft}_2}(N,\as)
= \bar g_0(\as) \,
\exp\Bigg\{\frac{1}{2\pi i} \int_0^{\frac{W}{2\beta_0\as}} dw\, e^{-w} \oint \frac{d\xi}{\xi} \,\Sud\(\as, -\frac{w}{\xi}\)
\\ \times 
\[\AP2\left[ \frac{\Gamma(N-\xi/2)}{\Gamma(N+\xi/2)}\]-\frac1{\Gamma(1+\xi)}\]\Bigg\},
\label{eq:A-soft2-BP}
\end{multline}
where $W$ is a cut-off, minimally set to $W=2$.
The inverse Mellin transform of the above result is finite, and can then be used to compute hadron-level
cross sections.

We have noticed in previous papers~\cite{Ball:2013bra,Bonvini:2014jma}
that one of interesting consequences of using
$A\text{-soft}_2$ instead of $N$-soft is a better perturbative
behavior, which is partially due to the resulting constant multiplying
the resummation, $\bar g_0$ or $g_0$ respectively. In fact, while the
perturbative expansion of the function $g_0$ is known to be poor,
driving large fixed-order corrections to the Higgs production, the
function $\bar g_0$ has a much more stable perturbative
expansion~\cite{Ball:2013bra, Bonvini:2014jma}. The relation between
the two constants is given in Eq.~\eqref{eq:g0bardef}, from which it
is clear that by using $\bar g_0$, we are effectively exponentiating
and, hence, resumming, part of the constant contribution. This is
similar in spirit to the so-called $\pi^2$-resummation discussed in
Refs.~\cite{PI_SQ, Eynck:2003fn} for Higgs and Drell-Yan processes.

We can also go a step further and try to exponentiate
the whole constant term~\cite{Eynck:2003fn}. To N$^3$LL
accuracy this amounts to replacing in the resummed expressions
$\bar g_0(\as)$ with
\beq \label{eq:expg0bar}
\bar G_0(\as)= \exp \[\as \bar g_{0,1}+\as^2\(\bar g_{0,2}-\frac{\bar g_{0,1}^2}{2}\) +\as^3 \(\bar g_{0,3}-\bar g_{0,1}\bar g_{0,2}+\frac{\bar g_{0,1}^3}{3}  \)+\Ord\(\as^4\)\].
\eeq
By construction the difference between $\bar g_0$ and $\bar G_0$ is $\Ord\(\as^4\)$
and hence beyond N$^3$LL accuracy considered here.
Because of the good convergence of the perturbative
expansion of $\bar g_0$ we do not expect the result obtained with
$\bar G_0$ to be much different compared to the one obtained with
$\bar g_0$. We will further comment on this in Sect.~\ref{sec:results},
where we present our hadron-level results.

\subsection{$\psi$-soft$_2$}\label{sec:psi-soft}

Before turning our attention to phenomenology, we discuss an alternative
option for the soft terms, which is very similar to the one used in our improved
result $A$-soft$_2$.
We first write $\Dh_k(N)$ in Eq.~\eqref{eq:Dhat} as~\cite{Bonvini:2012sh}
\begin{align}
\Dh_k(N)
&= \frac{1}{k+1}\sum_{j=0}^{k+1}\binom{k+1}{j}\, \Gamma^{(j)}(1)\, \[-\psi_0(N)\]^k \times \[1+\Ord\(\frac1{N^2}\)\] \nonumber\\
&= \frac{\Gamma^{(k+1)}(1)}{k+1} + \Dm_k\(\exp\psi_0(N)\) \times\[1+\Ord\(\frac1{N^2}\)\],
\end{align}
where $\psi_0(N)=\Gamma'(N)/\Gamma(N)$ is the DiGamma function.
Thus, except for the constant term, the functions $\Dh_k(N)$
are equivalent to the corresponding $\Dm_k(N)$ after the replacement
$\log N \to \psi_0(N)$, up to corrections of order $1/N^2$.
This means that Eq.~\eqref{eq:Cres2} can be upgraded by simply replacing $\log N$ with
$\psi_0(N)$, thereby restoring the analyticity properties of the coefficient function.

Therefore, we propose a new prescription, called $\psi$-soft$_2$, where
we also include Altarelli-Parisi improvement,
\begin{align}\label{eq:psi-soft2}
C_{\psi\text{-soft}_2}(N,\as)
&= g_0(\as) \,\exp \Big\{\AP2\[ \Sud\(\as,\psi_0(N)\) \] \Big\}\,,
\end{align}
which has the advantage of having almost all the good properties captured by $A$-soft$_2$, while being numerically very fast, because it can be computed using the MP.
In fact, any existing soft-gluon resummation code can be easily upgraded 
to $\psi$-soft$_2$.

However, Eq.~\eqref{eq:psi-soft2} has a clear disadvantage with respect to
$A$-soft$_2$, Eq.~\eqref{eq:A-soft2-BP}, namely the presence of the function $g_0$
rather than $\bar g_0$.
We have already commented in Sect.~\ref{sec:A2-soft} about the poor perturbative behavior
of $g_0$ compared to that of $\bar g_0$.
Here we limit ourselves to mention that, in this case, writing $g_0$ in exponential
form
\beq \label{eq:expg0}
G_0(\as)= \exp \[\as g_{0,1}+\as^2\(g_{0,2}-\frac{g_{0,1}^2}{2}\) +\as^3 \(g_{0,3}-g_{0,1}g_{0,2}+\frac{g_{0,1}^3}{3}  \)+\Ord\(\as^4\)\],
\eeq
as done for $\bar g_0$ in Eq.~\eqref{eq:expg0bar}, can improve significantly
the perturbative stability of the resummed result.
We anticipate that a phenomenological study shows that, after exponentiation
of both $g_0$ and $\bar g_0$, results obtained with $A$-soft$_2$ and $\psi$-soft$_2$
are very similar.

\section{Hadron-level results}\label{sec:results}

In this section we present numerical results for the Higgs production cross
section in the gluon fusion channel at the LHC. Our result correctly
accounts for finite top mass ($\mt=172.5$~GeV) at
NNLO~\cite{higgsuptoNNLO-finite-mt} and NNLL.
We also include bottom and charm masses in the LO prefactor (thus changing the overall normalization)
and in the NLO contributions, using the results of Ref.~\cite{Bonciani:2007ex}. The dependence on bottom and charm masses in the resummation is NLL accurate.
For consistency with the NNLO set of parton distribution functions NNPDF2.3~\cite{NNPDF23} with $\as(\mz)=0.118$,
that we adopt for our phenomenological analysis, we use
$\mbottom=4.75$~GeV and $\mcharm=1.41$~GeV. 
We refer the reader to Ref.~\cite{Forte:2013mda} for a discussion about the role of N$^3$LO parton densities.
We use $\mh=125$~GeV.

\subsection{A study of the scale dependence}\label{sec:scales}

\begin{figure}[t]
  \centering
  \includegraphics[width=0.495\textwidth,page=1]{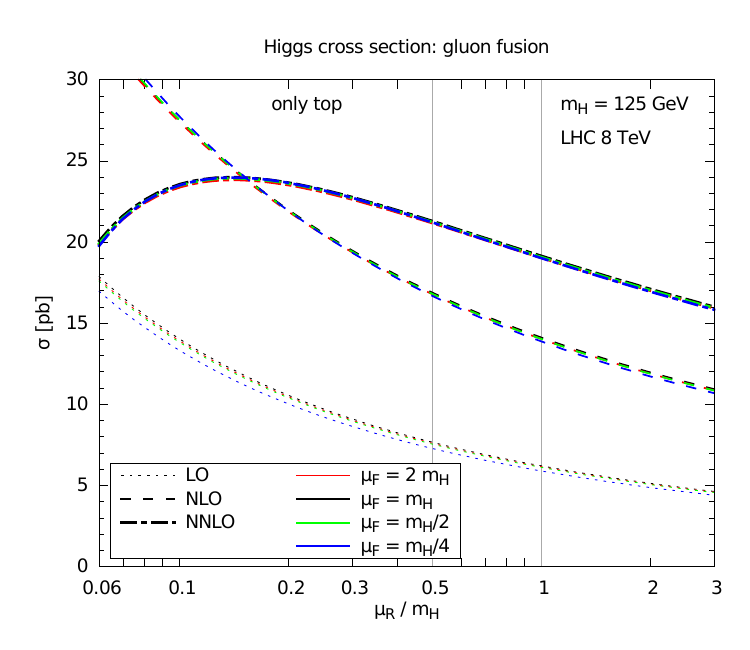}
  \includegraphics[width=0.495\textwidth,page=2]{images/paper_hadr_xsec_125_8.pdf}
  \caption{Renormalization scale dependence of LO, NLO and NNLO cross section.
    We show the effect of including just the top (left panel) and also bottom and charm (right panel)
    in the loop. The hadronic center of mass energy is $\sqrt{s}=8$~TeV.}
  \label{fig:8TeV_FO}
\end{figure}
In order to make contact with our previous work~\cite{Ball:2013bra}, we start by
showing in Fig.~\ref{fig:8TeV_FO} the cross section as a function of
$\mur$ at LO, NLO, and NNLO with only top included in the loop (left panel)
and with also bottom and charm (right panel).
The results have been computed using the code \texttt{\href{http://www.ge.infn.it/~bonvini/higgs/}{ggHiggs}}.
The collider energy is $\sqrt{s}=8$~TeV.
We use different colors for different values of $\muf$ in all curves,
for $\muf = \left\{2,1,1/2,1/4\right\}\mh$. We show four choices because the typical scale variation
is by a factor of two about its central value, but the central value is sometimes suggested
to be $\mh$ (e.g.~\cite{Catani:2003zt}) and sometimes $\mh/2$ (e.g.~\cite{Anastasiou:2012hx}).
It is interesting to observe that fixed-order results with only the top quark in the loop, left-hand plot of Fig.~\ref{fig:8TeV_FO}, barely depend on $\muf$:
this is due to a compensation between different channels (mainly $gg$ and $qg$ channels)~\cite{Ball:2013bra}.
However, this cancellation is not as perfect at NNLO, when we introduce bottom and charm contributions, right-hand plot of Fig.~\ref{fig:8TeV_FO}. This happens because in our framework, these corrections are correctly implemented only at NLO and there are no $\mbottom$, $\mcharm$ dependent contributions at $\Ord \(\as^2\)$ to compensate the NLO $\muf$ dependence. We observe that the main effect of including bottom and charm in the loop is to significantly reduce the cross section.

\begin{figure}[t]
  \centering
  \includegraphics[width=0.495\textwidth,page=4]{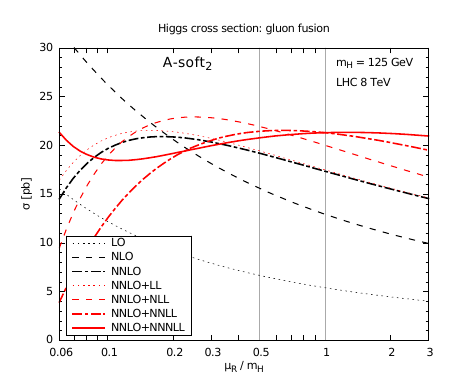}
  \includegraphics[width=0.495\textwidth,page=4]{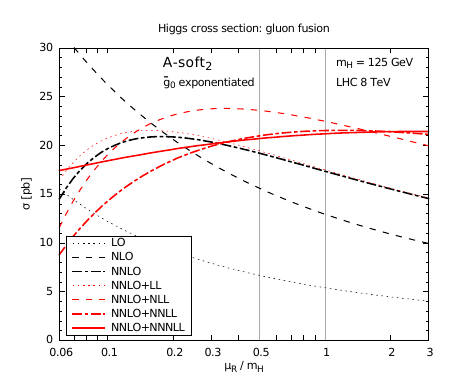}
  \caption{Our best prescription for the resummation, namely
    $A$-soft$_2$ described in Sect.~\ref{sec:A2-soft}, plotted as a
    function of the renormalization scale $\mur$. The factorization
    scale is $\muf=\mh$. We show fixed-order results as well as
    resummed ones. The plot on the left is obtained with the overall
    constant $\bar g_0$, while the one on the right with its
    exponentiated version $\bar G_0$, as defined in
    Eq.~\eqref{eq:expg0bar}.}
  \label{fig:8TeV_BP_NkLL}
\end{figure}

We now move to resummation.
In order to study the effect of different logarithmic orders,
we show  in Fig.~\ref{fig:8TeV_BP_NkLL} the resummation at LL, NLL, NNLL and N$^3$LL accuracy\footnote{%
We are adopting Notation*, see Table~\ref{tab:count}, so N$^3$LL is the currently highest possible accuracy.},
always matched to the same NNLO contribution, as a function of
$\mur$, for fixed $\muf=\mh$. We also show, for comparison, LO, NLO and NNLO curves.
The fixed order results have been computed using the code \texttt{\href{http://www.ge.infn.it/~bonvini/higgs/}{ggHiggs}},
while for the resummation we have written a new code called \texttt{\href{http://www.ge.infn.it/~bonvini/higgs/}{ResHiggs}}.
The plots show our best prediction, $A$-soft$_2$, with $\bar g_0$ (left panel) and its exponentiated version
$\bar G_0$ (right panel).
It is interesting to observe that exponentiating $\bar g_0$ leads to a flatter resummed result,
thereby suggesting that its exponentiation is probably improving the convergence of the series.
We also observe that, in any case, the N$^3$LL result is very similar in both cases over
a wide range of scales, so the exponentiation of $\bar g_0$ does not change significantly
the final result, as we have anticipated at the end of Sect.~\ref{sec:A2-soft}.
In both cases, we note that the inclusion of soft-gluon resummation at N$^3$LL significantly
reduces the $\mur$ scale uncertainty of fixed-order results and of previous resummed orders.

\begin{figure}[t]
  \centering
  \includegraphics[width=0.495\textwidth,page=1]{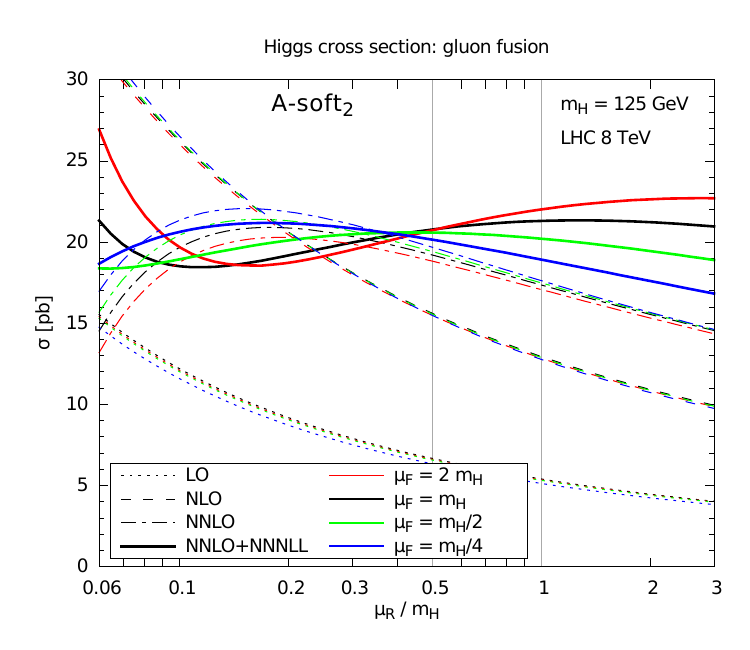}
  \includegraphics[width=0.495\textwidth,page=1]{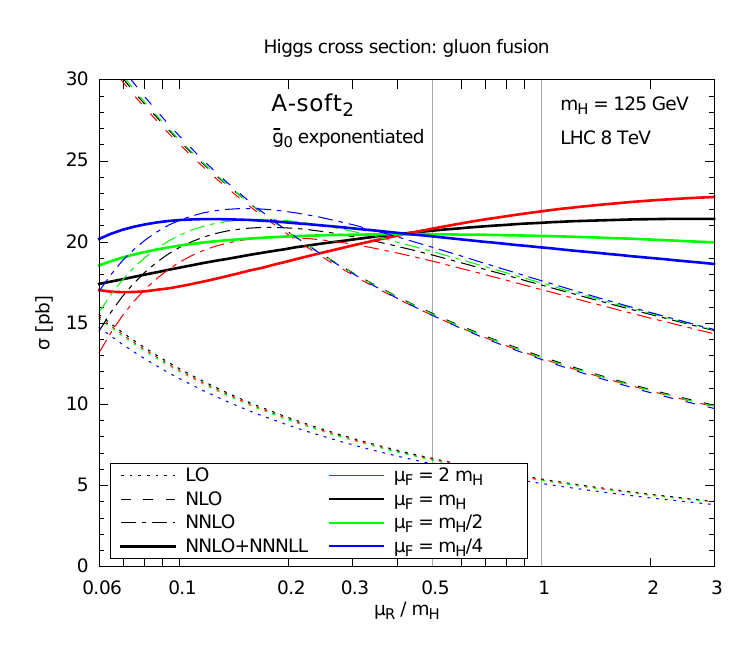}
  \caption{Our best result, namely NNLO+N$^3$LL with the $A$-soft$_2$
    resummation described in Sect.~\ref{sec:A2-soft}, plotted as a
    function of the renormalization scale $\mur$. Different colors
    correspond to different choices of the factorization scale is
    $\muf$. The plot on the left is obtained with the overall constant
    $\bar g_0$, while the one on the right with its exponentiated
    version $\bar G_0$, as defined in Eq.~\eqref{eq:expg0bar}.}
  \label{fig:8TeV_BP}
\end{figure}

In Fig.~\ref{fig:8TeV_BP} we concentrate on NNLO+N$^3$LL and also show the effect of varying $\muf$.
Since the resummation involves only the $gg$ channel, the resummed result depends more significantly
on the scale $\muf$, although formally such dependence is of order $\as^3$ with respect to the Born cross section.
Over a range of roughly a factor of $2$ about $\mur=\mh/2$ the results with (right panel) or without
(left panel) exponentiation of $\bar g_0$ are very similar, while they differ (and are more sensitive to $\muf$)
for more extreme choices of $\mur$ (especially at small $\mur$).
In these regions, the result obtained exponentiating $\bar g_0$ looks more
sensible and stable, suggesting, once again, that exponentiating $\bar g_0$ provides a
more stable result.
Moreover, we notice that NNLO+N$^3$LL result with $\muf= \mh/\,2$ barely depends on $\mur$.
We also observe that resummed curves for different values of $\muf$ approximately coincide
for a value of $\mur$ slightly smaller than $\mh/2$. 

\begin{figure}[t]
  \centering
  \includegraphics[width=0.495\textwidth,page=1]{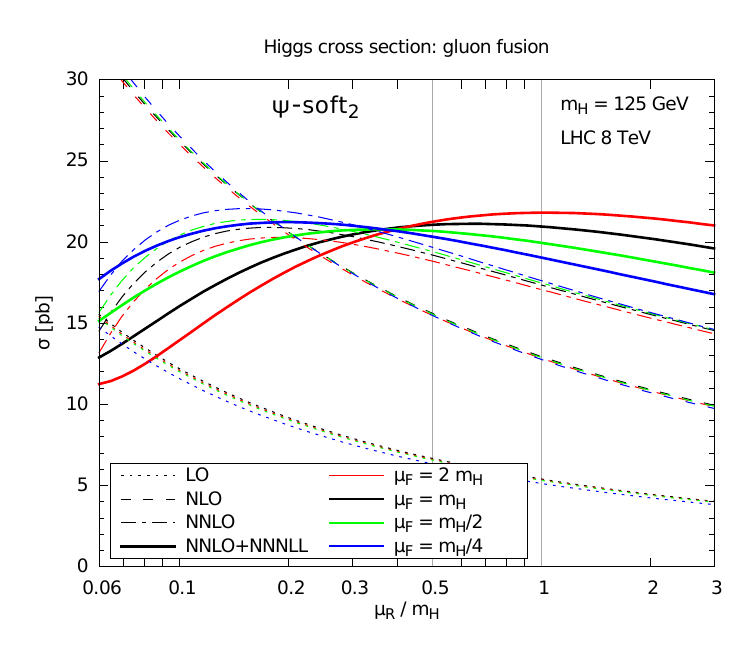}
  \includegraphics[width=0.495\textwidth,page=1]{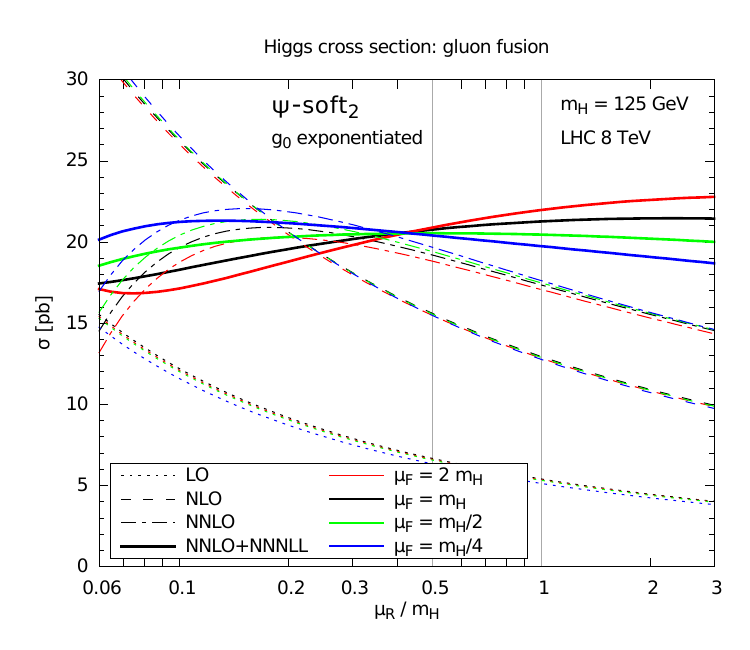}
  \caption{NNLO+N$^3$LL for $\psi$-soft$_2$ described in
    Sect.~\ref{sec:psi-soft}, plotted as a function of the
    renormalization scale $\mur$. Different colors correspond to
    different choices of the factorization scale is $\muf$. The plot
    on the left is obtained with the overall constant $g_0$, while the
    one on the right with its exponentiated version $G_0$.}
  \label{fig:8TeV_MP}
\end{figure}

In Fig.~\ref{fig:8TeV_MP} we show the same plots as in Fig.~\ref{fig:8TeV_BP}, but this time
obtained with the $\psi$-soft$_2$ prescription.
Since now the constant function in front of the exponential is $g_0$ rather than $\bar g_0$,
we can expect a result different from that of $A$-soft$_2$, when $g_0$ is not exponentiated (left panel).
However, the result with $G_0$ (right panel) is very similar to
the analogous result with $A$-soft$_2$.
It follows that $\psi$-soft$_2$ provides an acceptable alternative to our best choice $A$-soft$_2$,
provided that $G_0$ is used, i.e\ with $g_0$ exponentiated.
Since the numerical implementation of $\psi$-soft$_2$ is much faster than that of $A$-soft$_2$,
its usage can be convenient.

\begin{figure}[t]
  \centering
  \includegraphics[width=0.495\textwidth,page=4]{images/paper_hadr_xsec_res_125_8_MP00_AP0.pdf}
  \includegraphics[width=0.495\textwidth,page=1]{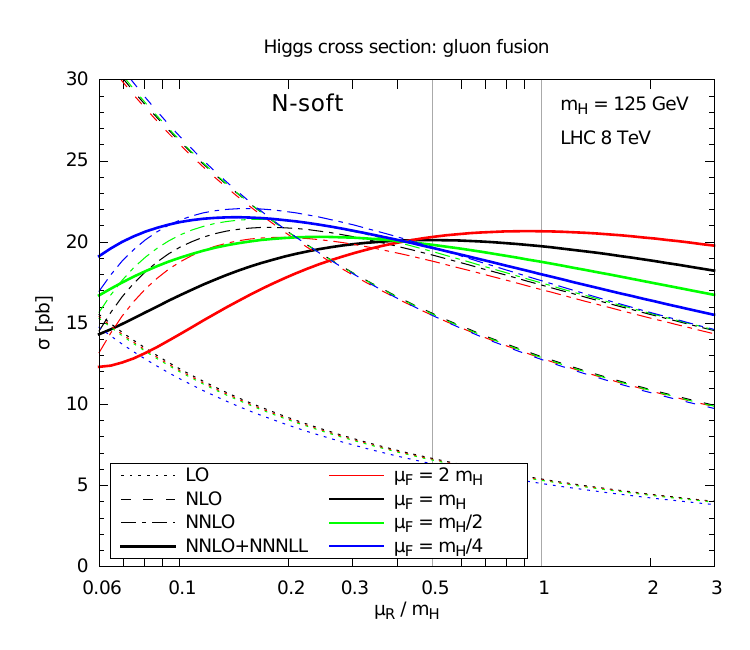}
  \caption{Result for $N$-soft resummation. On the left we show the
    resummation at different accuracies, always matched to the same
    NNLO result, for $\muf=\mh$, as a function of $\mur$. On the right
    we focus on the NNLO+N$^3$LL result and we also vary $\muf$.}
  \label{fig:8TeV_Nsoft}
\end{figure}

We show the result of the more traditional $N$-soft resummation in Fig.~\ref{fig:8TeV_Nsoft}. This is interesting because the value of the Higgs production cross section which is currently recommended by the Higgs Cross Section Working Group~\cite{hxswg} is based on Refs.~\cite{deFlorian:2012yg}, which includes $N$-soft resummation to NNLL accuracy.
In this case the difference between N$^3$LL and NNLL is bigger than it
was for $A$-soft$_2$, as it is shown on the left-hand plot of Fig.~\ref{fig:8TeV_Nsoft}.
This difference is partly due to the fact that $g_{0,3}$ is much larger than $\bar g_{0,3}$~\cite{Ball:2013bra,Bonvini:2014jma}.
The dependence of renormalization and factorization scale at N$^3$LL is comparable
$\psi$-soft$_2$ when $g_0$ is not exponentiated; however, the typical increase
of the cross section is smaller than what we find with our result.

\begin{figure}[t]
  \centering
  \includegraphics[width=\textwidth]{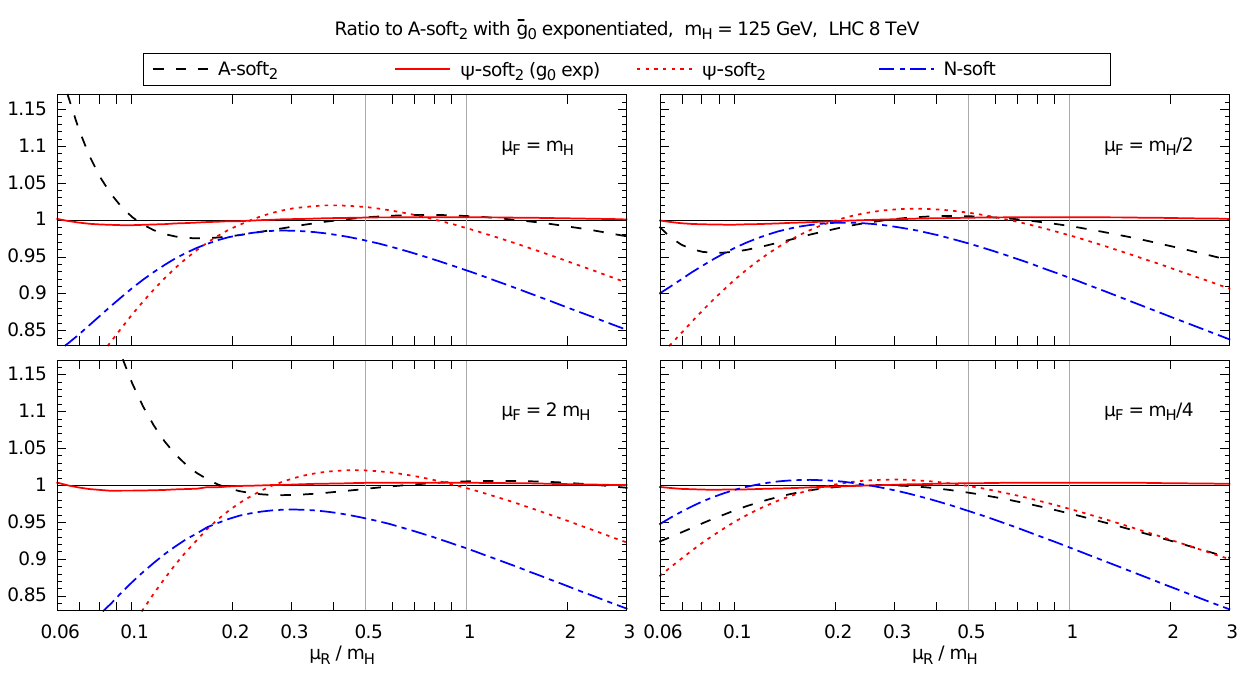}
  \caption{Ratios of different resummed results to our best prediction $A$-soft$_2$ with the exponentiated constant $\bar G_0$, plotted as a function of $\mur$, for different choices of $\muf$.}
  \label{fig:8TeV_ratios}
\end{figure}
A quantitative comparison between the different resummed results is shown in Fig.~\ref{fig:8TeV_ratios}, where ratios to our best prediction, namely $A$-soft$_2$ with the exponentiated constant $\bar G_0$, are plotted as a function of $\mur$, for different choices of
$\muf$. As previously observed, we confirm here quantitatively that the result obtained with $\psi$-soft$_2$ with $g_0$ exponentiated (solid red line) is almost identical to our best prediction, the difference being always below 1\%, and confirming that this prescription can be indeed used as a numerically convenient alternative to $A$-soft$_2$ with $\bar G_0$.
We also observe that for a wide choice of scales not exponentiating $\bar g_0$ in $A$-soft$_2$ (dashed black line) leads to a result which only differs from the result with $\bar G_0$ by a few percent. In contrast, the difference between resummed results with $g_0$, e.g.\ dotted red curve, or its exponentiated version $G_0$, e.g.\ solid red curve, is more pronounced. Finally, we note that the difference between our best prediction and $N$-soft is about 10\% for $\mur=\muf=\mh$.

\begin{figure}[t]
  \centering
  \includegraphics[width=0.495\textwidth,page=1]{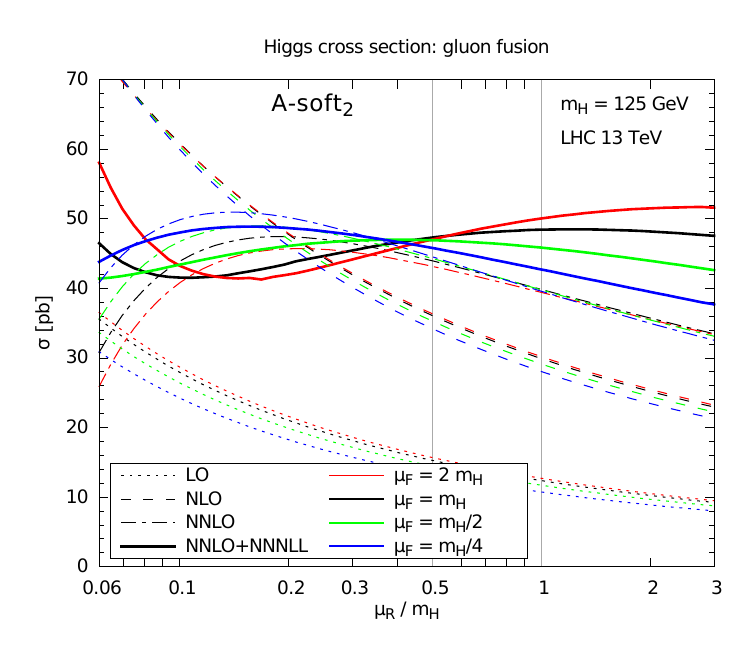}
  \includegraphics[width=0.495\textwidth,page=1]{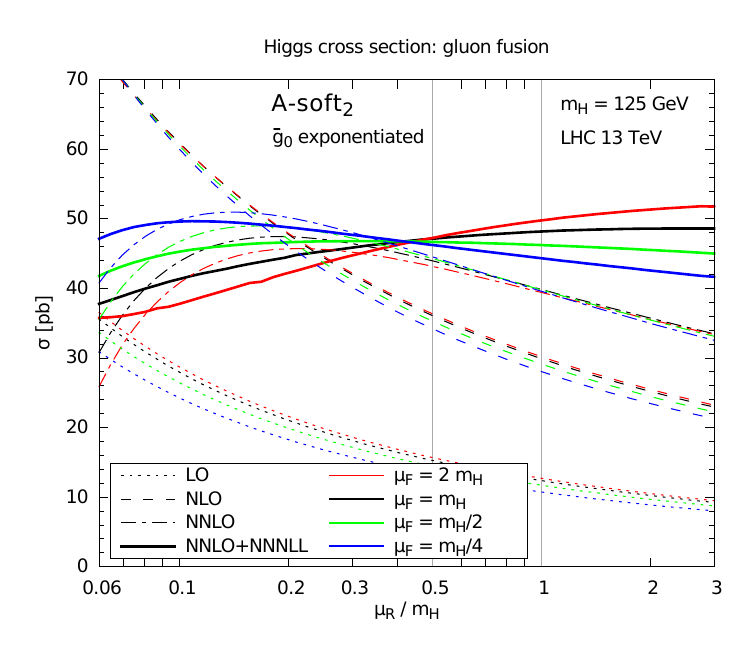}
  \caption{Same as Fig.~\ref{fig:8TeV_BP} but for $\sqrt{s}=13$~TeV.}
  \label{fig:13TeV_BP}
\end{figure}
\begin{figure}[t]
  \centering
  \includegraphics[width=0.495\textwidth,page=1]{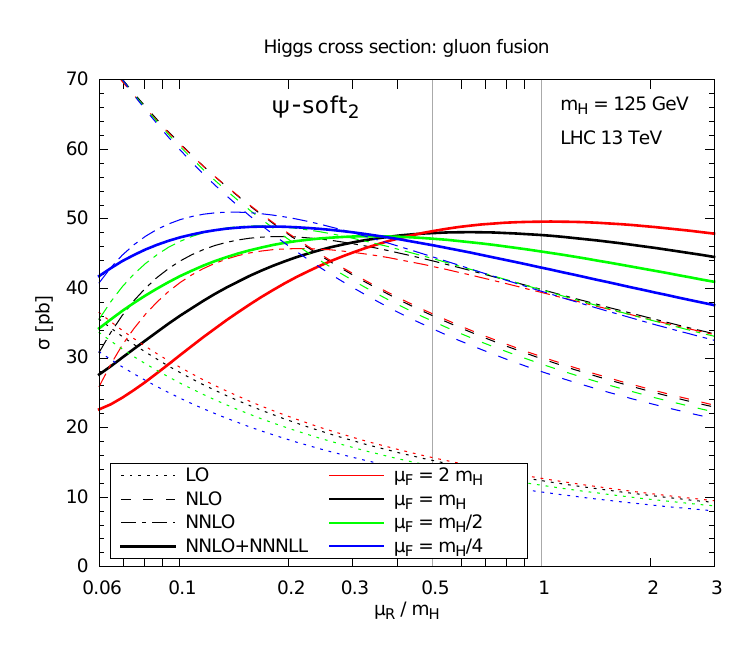}
  \includegraphics[width=0.495\textwidth,page=1]{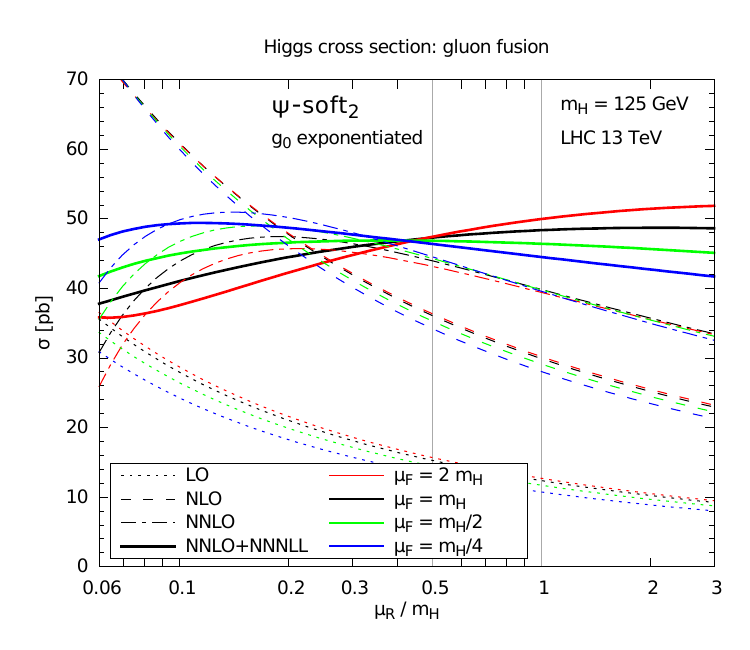}
  \caption{Same as Fig.~\ref{fig:8TeV_MP} but for $\sqrt{s}=13$~TeV.}
  \label{fig:13TeV_MP}
\end{figure}
In Figs.~\ref{fig:13TeV_BP} and \ref{fig:13TeV_MP}, we show the
results for $A$-soft$_2$ and $\psi$-soft$_2$ with collider energy
$\sqrt{s}=13$~TeV. The shape of the NNLO+N$^3$LL cross section in
terms of renormalization and factorization scales, is very similar to
the ones we have found at $8$~TeV, both for $A$-soft$_2$ and
$\psi$-soft$_2$, and the analogous ratios showed in Fig.~\ref{fig:8TeV_ratios}
for $8$~TeV look almost identical at $13$~TeV.
Thus, most of the points we have discussed for the
$8$~TeV case, also apply at $13$~TeV.
We just note that the NLO cross section exhibits a somewhat larger
$\muf$ dependence at $13$~TeV than at $8$~TeV.

\subsection{Numerical results for different collider energies}\label{sec:num}

In Table~\ref{tab:xsec} we summarise the results for our best
prediction, namely $A$-soft$_2$ with $\bar G_0$, for different
collider energies and different values of the Higgs mass. 
We estimate the theoretical uncertainty by
independently varying the scales up and down,  by a factor of two $m_\text{x}/2<\mur, \muf<2m_\text{x}$,
with the condition $1/2<\mur/ \muf<2$, where $m_\text{x}=\mh$ (on the left) and $m_\text{x}=\mh / 2$ (on the right).

Several comments can be made on the results in Table~\ref{tab:xsec}.
We first discuss on the effect of the resummation with respect to
NNLO.  If $\mur=\muf=\mh$ is chosen as the central scale, we find that
the inclusion of the resummation increases the NNLO prediction by 22\%
at $\sqrt{s}=7,8$~TeV and 21\% at $\sqrt{s}=13,14$~TeV. This is
consistent with the 16\% increase we previously found just for
N$^3$LO$_\text{approx}$~\cite{Ball:2013bra,Bonvini:2014jma}. We remind
the reader that the absolute value of the cross sections reported in
Table~\ref{tab:xsec} is lower than the one of our previous studies
because we now include bottom and charm quarks in the loops.
On the other hand, if $\mur=\muf=\mh/\,2$ is chosen as the central
scale, we find that the resummation only corrects the NNLO result by
roughly 5.5\% at $\sqrt{s}=7,8$~TeV and $\sqrt{s}=13,14$~TeV. 

\begin{table}[t]
  \centering
  \begin{tabular}{ccc}
    \multicolumn{3}{c}{$\mur=\muf=\mh$} \\[1.5ex]
    $\sqrt{s}$ & NNLO &  NNLO+N$^3$LL \\
    \midrule
    \phantom{0}7~TeV & $13.59^{+1.64}_{-1.46}$~pb & $16.65^{+1.42}_{-0.63}$~pb \\
    \phantom{0}8~TeV & $17.36^{+2.06}_{-1.81}$~pb & $21.19^{+1.81}_{-0.81}$~pb \\
    13~TeV           & $39.86^{+4.45}_{-3.79}$~pb & $48.19^{+4.15}_{-1.95}$~pb \\
    14~TeV           & $44.94^{+4.94}_{-4.20}$~pb & $54.27^{+4.70}_{-2.25}$~pb \\
    33~TeV          & $161.5^{+16.2}_{-14.4}$~pb & $193.4^{+17.2}_{-9.3}$~pb
  \end{tabular}
  \qquad
  \begin{tabular}{ccc}
    \multicolumn{3}{c}{$\mur=\muf=\mh/2$}\\[1.5ex]
    $\sqrt{s}$ & NNLO &  NNLO+N$^3$LL \\
    \midrule
    \phantom{0}7~TeV & $15.23^{+1.62}_{-1.64}$~pb & $16.08^{+0.57}_{-0.09}$~pb \\
    \phantom{0}8~TeV & $19.42^{+2.07}_{-2.06}$~pb & $20.48^{+0.71}_{-0.13}$~pb \\
    13~TeV           & $44.31^{+4.81}_{-4.50}$~pb & $46.70^{+1.52}_{-0.46}$~pb \\
    14~TeV           & $49.89^{+5.47}_{-5.08}$~pb & $52.56^{+1.81}_{-0.53}$~pb \\
    33~TeV           & $177.6^{+20.0}_{-18.9}$~pb & $187.6^{+7.6}_{-3.6}$~pb
  \end{tabular}
  \caption{Values of the NNLO and NNLO+N$^3$LL ($A$-soft$_2$ with $\bar G_0$) gluon fusion cross section
    for selected values of the collider energy.
    We use NNPDF2.3~\cite{NNPDF23} with $\mh=125$~GeV, $\mt=172.5$~GeV, $\mbottom=4.75$~GeV and $\mcharm=1.41$~GeV.
    The central value is for $\mur=\muf= \mh$ (on the left) and $\mur=\muf= \mh/\,2$ (on the right).
    As detailed in the text, we recommend to evaluate theoretical uncertainties by scale variation around
    $\mur=\muf=\mh $.
    Electro-weak corrections are not included.}
  \label{tab:xsec}
\end{table}

It is
important to note that, while the NNLO cross sections at
$\mur=\muf=\mh $ and $\mur=\muf=\mh/\,2$ differ by more than 10\%, the
NNLO+N$^3$LL is much more stable and only varies by 3\% (in the
opposite direction with respect to the change of the
fixed-order). Thus, as expected, the resummation of soft-enhanced
contributions reduces the theoretical uncertainty related to the
choice of the hard scale. 

However, while the NNLO+N$^3$LL central values in Table~\ref{tab:xsec}
are rather similar, their uncertainties differ significantly. The total
uncertainty band\footnote{We quote the total uncertainty band,
  because the uncertainties we find are fairly asymmetric.} at
$\mur=\muf=\mh$ is around $12$-$13$\%, while at $\mur=\muf=\mh /\,2$ is
$4$-$4.5$\%. This dramatic reduction of the scale dependence can be
understood thanks to the study we have performed in
Sect.~\ref{sec:scales}, where we have noticed, for instance, that
$A$-soft$_2$ (and $\psi$-soft$_2$) curves with $\muf=\mh/ \,2$ barely
depend on $\mur$. However, in Sect.~\ref{sec:scales} we have also
noticed that resummed curves for different $\muf$ all meet to a point
which is not too far away from $\mur=\mh/\,2$, resulting into an
artificially small scale dependence, which is probably not
representative of the true theoretical uncertainty.

Therefore, in order not to underestimate the uncertainty of the
NNLO+N$^3$LL result, we recommend to vary the scales around
$\mur=\muf=\mh $, which results into a fairly conservative $12$-$13$\%
uncertainty band, yet significantly smaller than the $20$-$22$\% band of the NNLO result.
One can also imagine implementing less canonical
scale choices. For instance, the plots in Sect.~\ref{sec:scales} seem
to suggest that varying the scales around $\muf=\mh/2$ and $\mur=\mh$
would lead to a less conservative, but still reliable, estimate of the
theoretical uncertainty.
 
We also compute the cross section for a future, very energetic run of
the LHC, with $\sqrt{s}=33$~TeV\footnote{Because of their slow
  numerical convergence, NNLO+N$^3$LL results for $\sqrt{s}=33$~TeV
  have been computed using $\psi$-soft$_2$ rather than the default
  $A$-soft$_2$. We have checked that in all the other cases this
  choice leads to differences at the permille level.}. The relations
between fixed-order and resummed results, and their uncertainty bands,
are qualitatively similar to the ones found at lower energies.
However, we note that at $\sqrt{s}=33$~TeV we may be becoming
sensitive to high-energy (BFKL) corrections, that we are not resumming
here. These corrections have negligible impact on the inclusive Higgs
cross section at current LHC energies, but can play a significant role
for future high-energy runs.

Our analysis only includes QCD corrections. Electro-Weak
contributions~\cite{electro-weak} can be also taken into account at
NLO and if one assumes that they factorize from QCD corrections, they
typically lead to a few percent increase of the cross
section. Moreover, our calculations assume that the Higgs boson is
produced on its mass shell. However, off-shell effects can be relevant
for precision Higgs physics, even if the Higgs is
light~\cite{Kauer:2012hd}. We leave the inclusion of these effects, as well as a study of PDF uncertainties, to
future work.

Finally, it is interesting to compare our result to the lower-order prediction
NNLO+NNLL obtained with $N$-soft, which is the accuracy of the QCD
calculation of the cross section currently recommended by the
Higgs Cross Section Working Group~\cite{hxswg},
for the central scale $\muf=\mur=\mh$.
At $8$~TeV, for instance, the cross section we obtain\footnote{The number presented here
 slightly differs to the one in Ref.~\cite{hxswg}.
Most of the difference comes from the presence of Electro-Weak contributions
in the result of Ref.~\cite{deFlorian:2012yg}, which leads to an increase of
about $5$\%~\cite{electro-weak}.
The remaining difference (about $1$\%) is due to the different PDF set used and the lack of
finite top mass correction at NNLO in Ref.~\cite{deFlorian:2012yg}.}
at this accuracy is $18.57^{+2.06}_{-1.81}$~pb,
with a total uncertainty band of $15$\%.
Hence, our calculation leads to a $14\%$ increase with respect to what
is currently used by the LHC experiments, and reduces the theory uncertainty.

\section{Conclusions}

Accurate theoretical predictions for Higgs production in gluon-gluon
fusion are of primary importance for the LHC physics programme. In
this paper we have studied the all-order resummation of enhanced
contributions due to the emissions of soft gluons.

Our resummation ($A$-soft$_2$) differs from the traditional one
(called here $N$-soft), in that it is build in such a way to respect the
analyticity properties of coefficient functions~\cite{Ball:2013bra,
  Bonvini:2014jma}. We improve on our previous work by computing
all-order N$^3$LL matched to NNLO cross sections for proton-proton
collisions at different center-of-mass energies. 
Our result therefore accounts for all the contributions also present in the N$^3$LO soft-virtual approximation of Ref.~\cite{Anastasiou:2014vaa}. 
Matching to full N$^3$LO, when this becomes available, will be straightforward.

Because of its analyticity properties, our results can be also easily matched to the
resummation of high-energy (BFKL) contributions, which we leave for
future work. 
However, high-energy resummation by itself has a relatively small
direct numerical impact, at current LHC energies, so we do not expect
this inclusion to significantly alter the results presented in this
paper. This situation can change, if future, very high-energy colliders are considered.

We have also shown that the traditional resummation formula can be minimally modified
($\psi$-soft$_2$) in such a way that the final result is almost identical to $A$-soft$_2$,
when constants are exponentiated in both cases. This provides a fast alternative
to implement our improved soft-gluon resummation results.

The resummation presented here has been implemented in a code called
\texttt{\href{http://www.ge.infn.it/~bonvini/higgs/}{ResHiggs}}
which can be interfaced with the fixed-order code 
\texttt{\href{http://www.ge.infn.it/~bonvini/higgs/}{ggHiggs}}.
Both codes are publicly available at the website
\texttt{\href{http://www.ge.infn.it/~bonvini/higgs/}{http://www.ge.infn.it/$\sim$bonvini/higgs/}}.

We have shown that, at energies relevant for LHC Run-I and Run-II, the
resummation ($A$-soft$_2$) corrects the NNLO result by as much as 20\%
at $\mur=\muf=\mh$, while the correction is much smaller, 5.5\%, at
$\mur=\muf=\mh /\,2$.  
However, the central value of NNLO+N$^3$LL results depends very mildly on the scale choice.
Moreover, the result obtained at NNLO+N$^3$LL ($A$-soft$_2$) leads to
a $14$\% increase with respect to the $N$-soft result at NNLO+NNLL~\cite{deFlorian:2012yg}
at $\mur=\muf=\mh$,
corresponding to the accuracy recommended by the Higgs Cross Section Working Group~\cite{hxswg}
and currently used by the LHC experiments.
We have also argued that theoretical uncertainties are better
estimated by scale variations about $\mur=\muf=\mh$, with a resulting
$12$-$13$\% total band.

Our results have a moderate dependence on the renormalization
scale, which instead drives the theoretical uncertainty of fixed-order
calculations. The dependence on the factorization scale instead is
somewhat larger than in fixed-order calculations. This is mainly due
to the fact that we only resum the $gg$ channel. This situation is
likely to improve, if we were able to resum at least the leading logarithms in the $qg$
channel, which are of the form $\as^k \log^{2 k-1}(1-z)$, using
techniques similar to the ones developed in Ref.~\cite{qg-dis} for
deep-inelastic scattering or by resumming the whole class of the
so-called next-to-eikonal contributions~\cite{next-eikonal}.

\section*{Acknowledgments}
We thank Richard Ball, Stefano Forte and Giovanni Ridolfi for many useful discussions,
and Frank Tackmann for a critical reading of the manuscript.
The work of SM is supported by the UK's STFC.

\appendix
\section{Prescriptions and their implementations}
\label{app:prescriptions}

\subsection{Minimal Prescription}
\label{app:MP}

The physical cross section, Eq.~\eqref{eq:xs}, can be written in terms of the $N$-space
coefficient function as
\beq\label{eq:sigma_inverse_mellin}
\frac{\sigma(\tau,M^2)}{\tau\sigma_0(M^2)} = \frac1{2\pi i}
\int_{c-i\infty}^{c+i\infty} dN \, \tau^{-N}\,\Lum(N)\, C(N,\as),
\eeq
where we have suppressed the flavor indices.
The parameter $c$ has to be larger
than the real part of the rightmost singularity of the integrand.
At the resummed level, the presence of the cut for real $N>N_L=\exp\frac1{2\beta_0\as}$
makes it impossible to find such a value of $c$.

The Minimal Prescription consists in computing the integral Eq.~\eqref{eq:sigma_inverse_mellin}
at the resummed level by simply choosing $c$ to the left of the cut and to the right of all the other
singularities; moreover, the integration contour is rotated counterclockwise on the
upper plane $\Im N>0$ and clockwise on the lower plane $\Im N<0$ to guarantee
numerical convergence~\cite{Catani:1996yz}. We write
\beq\label{eq:MP}
\frac{\sigma(\tau,M^2)}{\tau\sigma_0(M^2)} \overset{\rm MP}{=} \frac1{2\pi i}
\int_{\rm MPc} dN \, \tau^{-N}\,\Lum(N)\, C_{\rm res}(N,\as),
\eeq
where by MPc we denote the contour described above.
The integral in Eq.~\eqref{eq:MP} is finite, and it is proven to be an asymptotic sum
of the divergent series of the order-by-order inverse Mellin transform of $\CNsoft$~\cite{Catani:1996yz}.
However, because of the presence of the cut, the integration cannot be closed to the right,
with the consequence that the result does not vanish for $\tau>1$.
If one tries to interpret the result as a convolution of a parton luminosity and a partonic cross section,
the latter does not vanish for $z>1$; this contribution from the unphysical region $z>1$
violates factorization, but it is exponentially suppressed in $\LQCD$~\cite{Catani:1996yz}.

From a practical point of view, the computation of the integral in Eq.~\eqref{eq:MP}
requires the knowledge of the parton luminosity $\Lum(N)$ for values of $N$ along the contour.
However, the Mellin transform
\beq\label{eq:LumN}
\Lum(N) = \int_0^1 dx\, x^{N-1}\, \Lum(x)
\eeq
converges numerically only for $\Re N>0$, while the MPc contour involves values of $N$ with negative real part.
Therefore, for an efficient practical realization of the MP, the Mellin transform
Eq.~\eqref{eq:LumN} must be computed analytically, meaning that we need a functional form
for $\Lum(x)$.

A very efficient way of approximating $\Lum(x)$ is obtained expanding on a basis of
Chebyshev polynomials the function
\beq
f(u) = \[\frac{x^\beta}{(1-x)^\gamma}\Lum(x)\]_{x=e^u}
\eeq
on the range $u_{\rm min}<u<0$, with $u_{\rm min}\leq \log\tau$, and where $\beta$ and $\gamma$ are parameters
aimed to make $f(u)$ as flat as possible (by default we use $\beta=1$ and $\gamma=0$).
Fast routines are available for computing the coefficient of the expansion
of $f(u)$ on a basis of Chebyshev polynomials.
After straightforward manipulations~\cite{Bonvini:2012sh} we are able to write,
for integer $\gamma$,
\beq
\Lum(x) = (1-x)^\gamma x^{-\beta} \sum_{k=0}^n c_k \log^k x
= \sum_{j=0}^\gamma \binom{\gamma}{j} (-1)^j x^{j-\beta} \sum_{k=0}^n c_k \log^k x,
\eeq
where $n$ is the order of the Chebyshev approximation and $c_k$ are coefficients
which depend on $\muf$.
Its Mellin transform is
\beq\label{eq:LumNapprox}
\Lum(N) = \sum_{j=0}^\gamma \binom{\gamma}{j} \sum_{k=0}^n c_k \frac{(-1)^{k+j} k!}{(N+j-\beta)^{k+1}},
\eeq
suggesting that a small value for $\gamma$ is advisable to reduce the number of terms in the sum.
This luminosity can then be used in Eq.~\eqref{eq:MP} for numerical evaluation.
The whole procedure leads to a very fast numerical implementation: therefore,
where the MP is applicable, its usage is preferred.

\subsection{Borel Prescription}
\label{app:BP}

A prescription that deals directly with the divergent nature
of the series of the order-by-order Mellin inversion of $\CNsoft$
has been proposed in Refs.~\cite{Forte:2006mi,Abbate:2007qv,Bonvini:2008ei,Bonvini:2010tp,Bonvini:2012sh}.
This prescription adopts a Borel method for summing the divergent series, and it is therefore called
Borel Prescription (BP). We briefly review here the derivation, while referring
to the original works for a detailed discussion.

The $k$-th derivative of a function can be written as
\begin{align}
  f^{(k)}(0)
  &= \frac{k!}{2\pi i} \oint \frac{d\xi}{\xi^{k+1}}\, f(\xi) \nonumber\\
  &= \frac{1}{2\pi i} \int_0^\infty dw\, e^{-w} \oint \frac{d\xi}{\xi}\, f(\xi) \,\(\frac{w}{\xi}\)^k,\label{eq:BPoperator}
\end{align}
where in the first line we have used the Cauchy formula (the integration encloses the singular point $\xi=0$)
and in the second line we have rewritten the $k!$ as an integral.
As a result, the $k$-th derivative of $f$ has been translated into an integral operator
acting on the $k$-th power of the variable $w/\xi$.

This operator can be used to translate a power of $\log N$ in Eq.~\eqref{eq:Cres2} into
\beq
\log^k\frac1N = \frac{1}{2\pi i} \int_0^\infty dw\, e^{-w} \oint \frac{d\xi}{\xi}\, N^{-\xi} \,\(\frac{w}{\xi}\)^k,
\eeq
so that we can rewrite
\beq
\CNsoft(N,\as) = \frac{1}{2\pi i} \int_0^\infty dw\, e^{-w} \oint \frac{d\xi}{\xi}\, N^{-\xi} \,\CNsoft(e^{-w/\xi},\as).
\eeq
Notice that the function $\CNsoft(e^{-w/\xi},\as)$ has a cut in $-2\beta_0\as w<\xi<0$, due to the Landau pole.
Therefore, the $\xi$ integration contour, which must encircle the cut, extends to minus infinity
for $w\to\infty$, where the oscillatory behavior of the integrand makes the integral divergent.
This shows that the series is not Borel summable.
The Borel Prescription is now formulated as~\cite{Forte:2006mi,Abbate:2007qv,Bonvini:2008ei,Bonvini:2010tp,Bonvini:2012sh}
\beq\label{eq:BP}
\CNsoft(N,\as) \overset{\rm BP}{=} \frac{1}{2\pi i} \int_0^{\frac{W}{2\beta_0\as}} dw\, e^{-w} \oint \frac{d\xi}{\xi}\, N^{-\xi} \,\CNsoft(e^{-w/\xi},\as),
\eeq
where the $w$ integral has been cut off, and $W$ is minimally $W=2$
(corresponding to the inclusion of twist $4$ terms).
This cut-off makes the integral convergent, because the cut extends always on a finite range.
Since the BP is applied directly to the coefficient function $\CNsoft$,
the physical cross section maintain its convolution structure, without any violation of factorization.

As stressed already in Sect.~\ref{sec:prescriptions}, the numerical result
of Eq.~\eqref{eq:BP} is virtually identical to that obtained with the MP Eq.~\eqref{eq:MP},
since for phenomenologically relevant kinematic configurations the series is behaving
perturbatively and the way the divergence of the series is tamed is immaterial~\cite{Bonvini:2012sh}.

One of the advantages of the BP is that the inverse Mellin transform can be computed
analytically, since the $N$ dependence is confined in the generating function $N^{-\xi}$.
The convolution with the parton luminosity can hence be computed directly in $z$ space,
without the need of any approximation (although in practice approximations might be convenient
numerically, see Ref.~\cite{Bonvini:2012sh}). The price to pay is a slower implementation,
since the BP consists of two integrals, while the MP needs just one integration.

Another advantage, which is the main reason to consider the BP, is the fact that the form of the
soft terms is completely under control.
Indeed, in the derivation of Eq.~\eqref{eq:Cres2} from Eq.~\eqref{eq:Cres}, a large-$N$ limit
of the Mellin transform of the soft terms Eq.~\eqref{eq:Dk} is taken, in order to transform
the complicated derivatives of $\Gamma$ functions into simple powers of $\log N$,
so that a closed form for the functions $g_i(\as\log N)$ can be found.
The BP makes possible to skip this large-$N$ limit, since the derivatives can be translated
into powers through the operator Eq.~\eqref{eq:BPoperator}.
This can be obtained by simply replacing the generating function $N^{-\xi}$ in Eq.~\eqref{eq:BP}
with the generating function of the derivatives in Eq.~\eqref{eq:Dhat}, which corresponds to
introducing soft terms with the right analyticity properties (see discussion there).

Moreover, in deriving Eq.~\eqref{eq:Cres2} all constant terms from the Mellin transform
of plus distributions have been (by choice) removed from the exponent and put into the function
$g_0(\as)$, Eq.~\eqref{eq:g0bardef}. Here we have the opportunity to also restore these constants
in the exponent, since they were there in the original expression Eq.~\eqref{eq:Cres}.

However, the proper way to restore the original logarithms and constants is to let
the BP operator act directly on the exponent $\Sud(\as,\log N)$.
Therefore, we propose a new version of the BP,
\beq\label{eq:BPexpg0exp}
C_{A\text{-soft}}(N,\as) \overset{\rm BP}{=} \bar g_0(\as)
\exp\Bigg\{\frac{1}{2\pi i} \int_0^{\frac{W}{2\beta_0\as}} dw\, e^{-w} \oint \frac{d\xi}{\xi}\,
\[\frac{\Gamma(N-\xi/2)}{\Gamma(N+\xi/2)} -\frac1{\Gamma(1+\xi)}\] \,\Sud\(\as, -\frac{w}{\xi}\)\Bigg\},
\eeq
which can be further improved by letting the Altarelli-Parisi operator, Eq.~\eqref{eq:AP-operator},
act on the generating function, as in Eq.~\eqref{eq:A-soft2-BP}.
Since the conversion of soft terms is done at the exponent, the two integrals
of the BP have to be computed for each value of $N$, and then the inverse Mellin
transform has to be computed numerically.
In practice, we can use the same numerical technique used for the MP
(an integral of the form Eq.~\eqref{eq:MP}, with the luminosity approximated as
Eq.~\eqref{eq:LumNapprox}), with the exception that now the function
Eq.~\eqref{eq:BPexpg0exp} does no longer have a cut,
and hence the inverse Mellin transform exists.
The resulting numerical implementation is slow, compared to the MP.
However, this procedure is, to our knowledge, the only viable solution to reproduce 
the structure of Eq.~\eqref{eq:Cres} to all orders in a numerical way.

\end{document}